\documentclass[prb,showpacs,amsmath,amssymb,showkeys]{revtex4}
\usepackage{amssymb}
\usepackage{amsmath}
\usepackage{epsf}
\usepackage[dvips]{epsfig}
\usepackage{graphicx}
\usepackage{dcolumn}
\usepackage{bm}

\begin{document}
\widetext
\title{Terahertz emitters and detectors based on carbon nanotubes}
\author{M.E.~Portnoi}
\email[]{m.e.portnoi@exeter.ac.uk} \affiliation{School of Physics,
University of Exeter, Stocker Road, Exeter EX4 4QL, United
Kingdom} \affiliation{International Center for Condensed Matter
Physics, University of Brasilia, 70904-970 Brasilia DF, Brazil}
\author{O.V.~Kibis}
\email[]{oleg.kibis@nstu.ru} \affiliation{Department of Applied
and Theoretical Physics, Novosibirsk State Technical University,
Novosibirsk 630092, Russia} \affiliation{International Center for
Condensed Matter Physics, University of Brasilia, 70904-970
Brasilia DF, Brazil}
\author{M.~Rosenau~da~Costa}
\email[]{rosenau@unb.br} \affiliation{International Center for 
Condensed Matter Physics, University of Brasilia, 70904-970 Brasilia DF, 
Brazil}
\date{28 August, 2006}
\begin{abstract}
We formulate and justify several proposals utilizing the unique
electronic properties of different types of carbon nanotubes for 
a broad range of applications to THz optoelectronics, including THz
generation by hot electrons in quasi-metallic nanotubes, frequency
multiplication in chiral-nanotube-based superlattices controlled
by a transverse electric field, and THz radiation detection and 
emission by armchair nanotubes in a strong magnetic field.
\end{abstract}
\maketitle
\section{Introduction}
Creating a compact reliable source of terahertz (THz) radiation is
one of the most formidable tasks of the contemporary applied
physics.\cite{revFerguson02} One of the latest trends in THz
technology\cite{revDragoman04} is to use carbon nanotubes ---
cylindrical molecules with nanometer diameter and micrometer
length\cite{Iijima91,SaitoBook98,DresselhausBook01,ReichBook04}
--- as building blocks of novel high-frequency devices.  There are
several promising proposals of using carbon nanotubes for THz
applications including a nanoklystron utilizing extremely
efficient high-field electron emission from
nanotubes,\cite{revDragoman04,6} devices based on negative
differential conductivity in large-diameter semiconducting
nanotubes,\cite{7,8} high-frequency resonant-tunneling
diodes\cite{9} and Schottky diodes,\cite{10,11,12,13} as well as
electric-field-controlled carbon nanotube superlattices,~\cite{14}
frequency multipliers,~\cite{15,16} THz amplifiers,~\cite{17}
switches~\cite{18} and antennas.~\cite{19}

In this Paper we will formulate and discuss several novel schemes
to utilize physical properties of single-wall carbon nanotubes
(SWNTs) for generation and detection of THz radiation.

\section{Quasi-metallic carbon nanotubes as terahertz emitters}

In this Section we propose a novel scheme for generating THz
radiation from single-walled carbon nanotubes (SWNTs). This scheme
is based on the electric-field induced heating of electron gas
resulting in the inversion of population of optically active
states with the energy difference within the THz spectrum range.
It is well-known that the elastic backscattering processes in
metallic SWNTs are strongly suppressed,~\cite{20} and in a high
enough electric field charge carriers can be accelerated up to the
energy allowing emission of optical/zone-boundary phonons. At this
energy, corresponding to the frequency of about 40 THz, the major
scattering mechanism switches on abruptly resulting in current
saturation.\cite{21,22,23,Freitag04,24} In what follows we show
that for certain types of carbon nanotubes the heating of
electrons to the energies below the phonon-emission threshold
results in the spontaneous THz emission with the peak frequency
controlled by an applied voltage.

\begin{figure}[htb]
\begin{center}
\includegraphics[width=6.0cm]{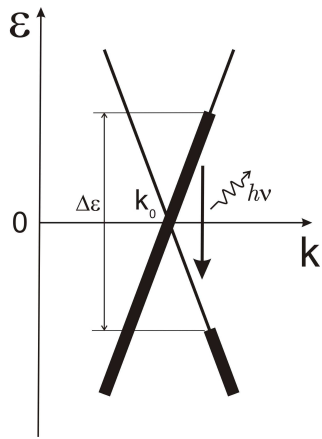}
\end{center}
\caption{The scheme of THz photon generation by hot carriers in
metallic SWNTs.} 
\label{fig1}
\end{figure}

The electron energy spectrum of metallic SWNTs $\varepsilon(k)$
linearly depends on the electron wave vector $k$ close to the
Fermi energy and has the form $\varepsilon(k)=\pm \hbar
v_F|k-k_0|$, where $v_F\approx 9.8\times 10^5$ m/s is the Fermi
velocity of graphene, which corresponds to the commonly used
tight-binding  matrix element $\gamma_0 =
3.033\;$eV.~\cite{SaitoBook98,DresselhausBook01,ReichBook04} Here
and in what follows the zero of energy is defined as the Fermi
energy position in the absence of an external field. When the
voltage, $V$, is applied between the SWNT ends, the electron
distribution is shifted in the way shown by the heavy lines in
Fig.~\ref{fig1} corresponding to the filled electron states.
This shift results in inversion of population and,
correspondingly, in optical transitions between filled states in
the conduction band and empty states in the valence band.  The
spectrum of optical transitions is determined by the distribution
function for hot carriers, which in turn depends on the applied
voltage and scattering processes in the SWNT. It is well-known
that the major scattering mechanism in SWNTs is due to
electron-phonon interaction.\cite{21,22,23,24} Since the
scattering processes erode the inversion of electron population,
an optimal condition for observing the discussed optical
transitions takes place when the length of the SWNT $L<l_{ac}$,
where the electron mean-free path for acoustic phonon scattering
is $l_{ac}\approx2.4\;\mu$m.\cite{23} Below we consider only such
short SWNTs with ideal Ohmic contacts\cite{22} and in the
ballistic transport regime, when the energy acquired by the
electron on the whole length of the tube, $\Delta\varepsilon=eV$,
does not exceed the value of $\hbar \Omega =0.16\;$eV at which the
fast emission of high-energy phonons begins.\cite{23} In this
regime,\cite{22,23} the current in the nanotube is given by the
B\"{u}ttiker-Landauer formula, $I\approx (4e^2/h)V$, and the
distribution function for hot electrons is
\begin{equation} f_e(k)=\left\{\begin{array}{rl}
1, &0<k-k_0<\Delta\varepsilon/2\hbar v_F\\
0, &k-k_0>\Delta\varepsilon/2\hbar v_F
\end{array}\right..
\label{fe}
\end{equation}
The distribution function for hot holes, $f_h(k)$, has the same
form as $f_e(k)$.

Let us select a SWNT with the crystal structure most suitable for
observation of the discussed effect. First, the required nanotube
should have metallic conductivity and, second, the optical
transitions between the lowest conduction subband and the top
valence subband should be allowed. The crystal structure of a SWNT
is described by two integers $(n,m)$, which completely define
their physical
properties.~\cite{Iijima91,SaitoBook98,DresselhausBook01,ReichBook04}
The SWNTs with true metallic energy band structure, for which the
energy gap is absent for any SWNT radius, are the armchair $(n,n)$
SWNTs only.\cite{ReichBook04,25,26,27,28} However, for armchair
SWNTs the optical transitions between the first conduction and
valence subbands are forbidden.\cite{Milosevic03,Jiang04} So we
propose to use for the observation of THz generation the so-called
quasi-metallic $(n,m)$ SWNTs with $n-m=3p$, where $p$ is an
non-zero integer. These nanotubes, which are gapless within the
frame of a simple zone-folding model of the $\pi$-electron
graphene spectrum,\cite{SaitoBook98} are in fact narrow-gap
semiconductors due to curvature effects. Their bandgap is given
by~\cite{25,28}
\begin{equation}
\varepsilon_g=\frac{\hbar v_F a_{\mbox{\scriptsize{C-C}}} 
\cos 3\theta}{8 R^2}, 
\label{Eg}
\end{equation}
where $a_{\mbox{\scriptsize{C-C}}}=1.42\;$\r{A} is the
nearest-neighbor distance between two carbon atoms, $R$ is the
nanotube radius, and $\theta=\arctan [\sqrt{3}m/(2n+m)]$ is the
chiral angle.~\cite{SaitoBook98} It can be seen from
Eq.~(\ref{Eg}) that the gap is decreasing rapidly with increasing
the nanotube radius.  For large values of $R$ this gap can be
neglected even in the case of moderate applied voltages due to
Zener tunneling of electrons across the gap.  It is easy to show
in the fashion similar to the original Zener's work\cite{Zener34}
that the tunneling probability in quasi-metallic SWNTs is given by
$\exp(-\alpha \varepsilon_g^2/eE \hbar v_F)$, where $\alpha$ is a
numerical factor close to unity.\cite{Note_alpha} For example, for
a zigzag $(30,0)$ SWNT the gap is $\varepsilon_g\approx 6$ meV and
the Zener breakdown takes place for the electric field $E\sim
10^{-1}$ V/$\mu$m, which corresponds to a typical voltage of 0.1 V
between the nanotube ends. In what follows all our calculations
are performed for a zigzag $(3p,0)$ SWNT of large enough radius
$R$ and for applied voltages exceeding the Zener breakdown, so
that the finite-gap effects can be neglected. The obtained results
can be easily generalized for any quasi-metallic large-radius
SWNT.

\begin{figure}
\begin{center}
\includegraphics[width=8.0cm]{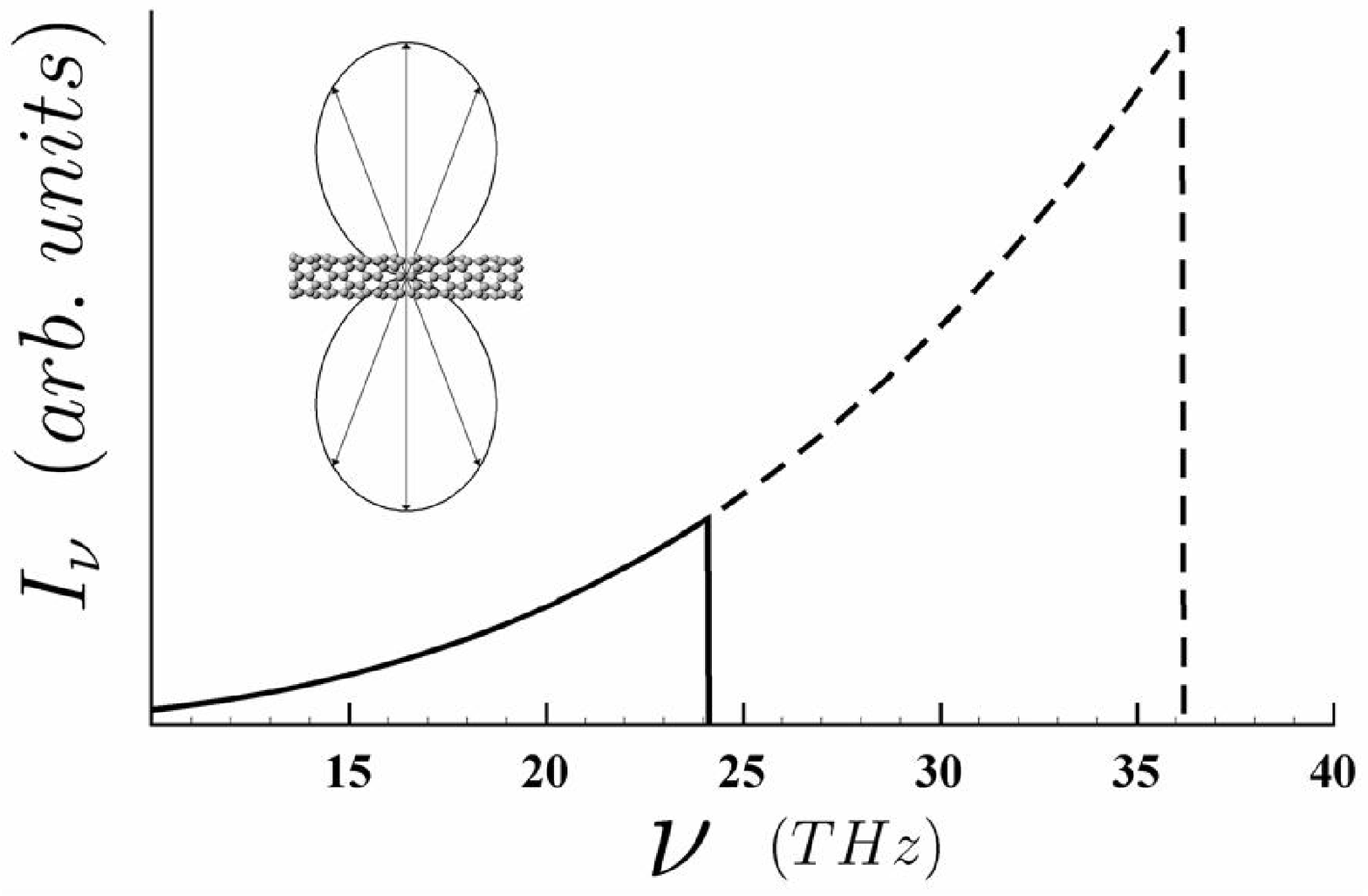}
\end{center}
\caption{The spectral density of spontaneous emission as a
function of frequency for two values of applied voltage: solid
line for $V=0.1\;$V; dashed line for $V=0.15\;$V. The inset shows
the directional radiation pattern of the THz emission with respect
to the nanotube axis.} \label{fig2}
\end{figure}

Optical transitions in SWNTs have been a subject of extensive
research (see, e.g., Refs.~\onlinecite{Milosevic03,Jiang04,
Grunes03,Popov04,Saito04,Goupalov05,Oyama06}; comprehensive 
reviews of the earlier work can be found in 
Refs.~\onlinecite{SaitoBook98,DresselhausBook01}). We treat 
these transitions using the results of the nearest-neighbor 
orthogonal $\pi$-electron tight binding model.~\cite{SaitoBook98} 
Despite its apparent simplicity and well-known limitations, this 
model has been extremely successful in describing low-energy optical 
spectra and electronic properties of single-walled SWNTs (see, e.g., 
Ref.~\onlinecite{Sfeir06} for one of the most recent manifestations 
of this model success).
Our goal is to calculate the spectral density of spontaneous
emission, $I_{\nu}$, which is the probability of optical
transitions per unit time for the photon frequencies in the
interval $(\nu,\;\nu + d\nu)$ divided by $d\nu$. In the dipole
approximation\cite{LandauQED} this spectral density is given by
\begin{equation}
I_\nu=\frac{8 \pi e^2 \nu}{3 c^3}\sum_{i,f} f_e(k_i) f_h(k_f)
\left| \left\langle\Psi_f\left|\hat{v}_z\right|\Psi_i\right\rangle
\right|^2 \delta(\varepsilon_i -\varepsilon_f- h\nu). \label{Wem}
\end{equation}
Equation~(\ref{Wem}) contains the matrix element of the electron
velocity operator. In the frame of the tight binding model, this
matrix element for optical transitions between the lowest
conduction and the highest valence subbands of the $(3p,0)$ zigzag
SWNT can be written as (cf. Ref.~\onlinecite{Jiang04,Grunes03})
\begin{equation}
\left\langle\Psi_f\left|\hat{v}_z\right|\Psi_i\right\rangle=
\frac{a_{\mbox{\scriptsize{C-C}}}\omega_{if }}{8}\delta_{k_f,k_i},
\label{dipole}
\end{equation}
where $\hbar\omega_{if}=\varepsilon_i-\varepsilon_f$ is the energy
difference between the initial $(i)$ and the final $(f)$  states.
These transitions are associated with the light polarized along
the nanotube axis $z$, in agreement with the general selection
rules for SWNTs.\cite{Milosevic03} Substituting Eq.~(\ref{dipole})
in Eq.~(\ref{Wem}) and performing necessary summation, we get
\begin{equation}
I_\nu=L f_e(\pi\nu/v_F) f_h(\pi\nu/v_F) \frac{\pi^2 e^2
a_{\mbox{\scriptsize{C-C}}}^2 \nu^3}{6 c^3 \hbar v_F}.
\label{Wfinal}
\end{equation}
Equation (\ref{Wfinal}) has broader applicability limits than the
considered case of $L<l_{ac}$ and  $eV<\hbar\Omega$, in which the
distribution functions for electrons and holes are given by
Eq.~(\ref{fe}). In the general case there is a strong dependence
of $I_\nu$ on the distribution functions, which have to be
calculated taking into account all the relevant scattering
mechanisms.\cite{21,22,23,24} In the discussed ballistic regime
the spectral density has a universal dependence on the applied
voltage and photon frequency for all quasi-metallic SWNTs. In
Fig.~\ref{fig2} the spectral density is shown for two values of
the voltage. It is clearly seen that the maximum of the spectral
density of emission has strong voltage dependence and lies in the
THz frequency range for experimentally attainable voltages.
The directional radiation pattern, shown in the inset of
Fig.~\ref{fig2}, reflects the fact that the emission of light
polarized normally to the nanotube axis is forbidden by the
selection rules for the optical transitions between the lowest
conduction subband and the top valence subband.

For some device applications it might be desirable to emit photons
propagating along the nanotube axis, which is possible in optical
transitions between the SWNT subbands characterized by angular
momenta differing by one.~\cite{ReichBook04,Milosevic03} To
achieve the emission of these photons by the electron heating, it
is necessary to have an intersection of such subbands within the
energy range accessible to electrons accelerated by attainable
voltages. From our analysis of different types of SWNTs, it
follows that the intersection is possible, e.g., for the lowest
conduction subbands in several semiconducting zigzag nanotubes and
in all armchair nanotubes. However, for an effective THz emission
from these nanotubes it is necessary to move the Fermi level very
close to the subband intersection point.\cite{KibisTPL05}
Therefore, obtaining the THz emission propagating along the
nanotube axis is a much more difficult technological problem
comparing to the generation of the emission shown in
Fig.~\ref{fig2}.

\section{Chiral carbon nanotubes as frequency multipliers}

\begin{figure}[htb]
\begin{center}
\includegraphics[width=8.0cm]{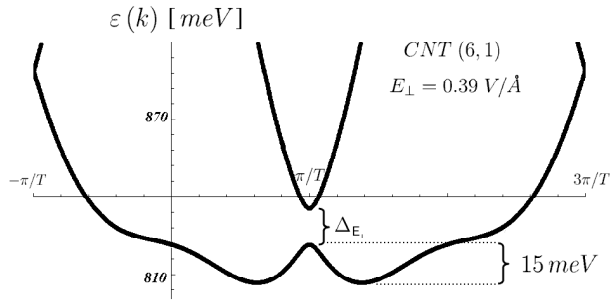}
\end{center}
\caption{Energy spectrum of the $(6,1)$ SWNT in a transverse
electric field, $E_{\perp}=4$ V/nm.} \label{fig3}
\end{figure}
Chiral nanotubes are natural superlattices. For example, a
$(10,9)$ single-wall nanotube has a radius which differs from the
radius of the most commonly studied $(10,10)$ nanotube by less
than five per cent, whereas a translational period along  the axis
of the $(10,9)$ SWNT is almost thirty times larger than the period
of the $(10,10)$ nanotube. Correspondingly, the first Brillouin
zone of the $(10,9)$ nanotube is almost thirty times smaller than
the first zone for the $(10,10)$ tube.  However such a Brillouin
zone reduction cannot influence electronic transport unless there
is a gap opening between the energy subbands resulting from the
folding of graphene spectrum. In our research we show how an
electric field normal to the nanotube axis opens noticeable gaps
at the edge of the reduced Brillouin zone, thus turning a
long-period nanotube of certain chirality into a `real'
superlattice. The field-induced gaps are most pronounced in
$(n,1)$ SWNTs.\cite{14,KibisTBP} Figure~\ref{fig3} shows the opening of
electric-field induced gap near the edge of the Brillouin zone of
a $(6,1)$ SWNT.
\begin{figure}[htb]
\begin{center}
\includegraphics[width=10.0cm]{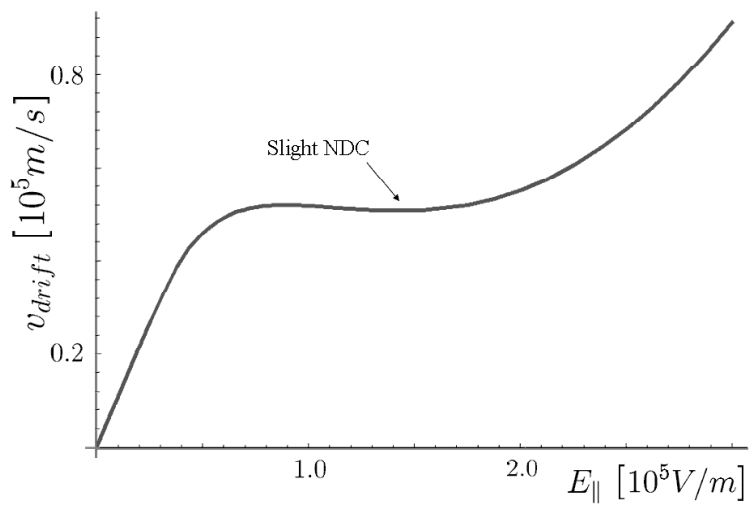}
\end{center}
\caption{The electron drift velocity in the lowest conduction
subband of a $\left( 6,1\right)$ SWNT as a function of the
longitudinal electric field, in the presence of acoustic phonons.}
\label{fig4}
\end{figure}
This gap opening results in the appearance of a
negative effective-mass region in the nanotube energy spectrum.
The typical electron energy in this part of the spectrum of 15 meV
is well below the optical phonon energy $\hbar\Omega\approx 160$
meV, so that it can be easily accessed in moderate heating
electric fields. The negative effective mass results in the
negative differential conductivity in a wide range of applied
voltages, as can be seen from Fig.~\ref{fig4}.

\begin{figure}[htb]
\begin{center}
\includegraphics[width=16.0cm]{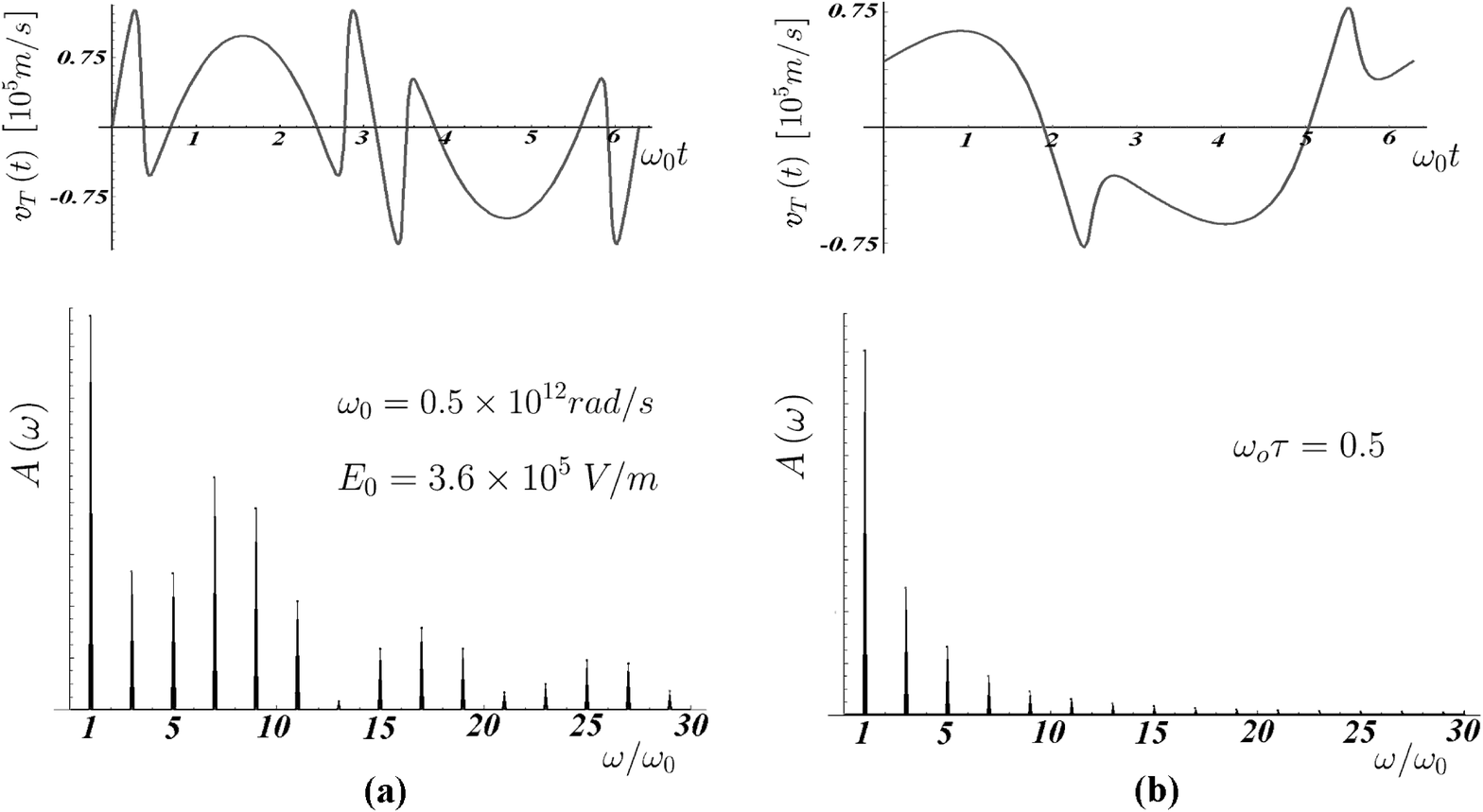}
\end{center}
\caption{Time dependence of the electron velocity in the lowest
conduction subband of a $\left( 6,1\right)$ SWNT under the
influence of a pump harmonic longitudinal electric field,
$E_{\parallel }\left( t\right) =E_{0}\sin \left( \omega
_{0}t\right) $, and its correspondent spectral distribution
$A\left( \omega \right)$: (a) in the ballistic transport regime;
(b) in the presence of scattering with the relaxation time
$\tau=10^{-12}s$.} \label{fig5}
\end{figure}

The effect of the negative effective mass also leads to an
efficient frequency multiplication in the THz range. The results
of our calculations of the electron velocity in the presence of
the time dependent longitudinal electric field are presented in
Fig.~\ref{fig5}.
One of the advantages of a frequency multiplier based on chiral
SWNTs, in comparison with the conventional
superlattices,\cite{Alek} is that the dispersion relation in our
system can be controlled by the transverse electric field
$E_\perp$.

\section{Armchair nanotubes in a magnetic field as tunable 
THz detectors and emitters}

\begin{figure}[htb]
\begin{center}
\includegraphics[width=11.0cm]{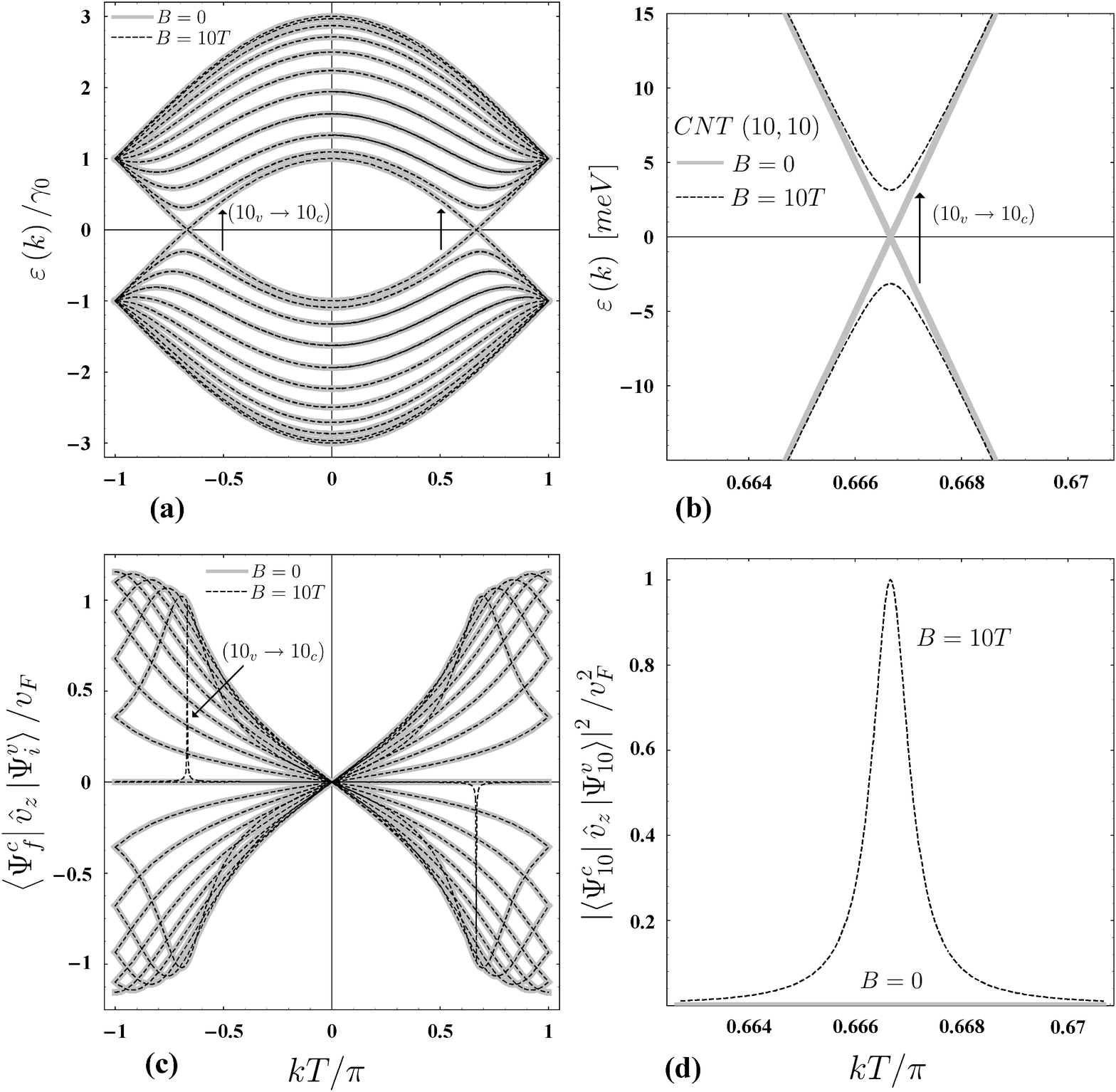}
\end{center}
\caption{
(a)~Band structure of a $\left( 10,10\right) $ nanotube, with and 
without an external magnetic field along the nanotube axis. 
(b)~Detailed view of the gap, which is opened between the top valence 
subband and the lowest conduction subband in an external field $B=10\;$T. 
(c)~The change in the dipole optical transitions matrix elements, 
for the light polarized along the SWNT axis, due to the introduction of 
the external magnetic field. The only appreciable change is in the 
appearance of a high narrow peak associated with the transition 
$\left( 10_{v}\rightarrow 10_{c}\right)$, which is not allowed in the 
absence of the magnetic field. 
(d)~Dependence of the squared dipole matrix element for the transition  
$\left(10_{v}\rightarrow 10_{c}\right)$ on the 1D wave vector $k$, 
with and without an external magnetic field.}
\label{fig6}
\end{figure}
The problem of detecting THz radiation is known to be at least as
challenging as creating reliable THz sources. Our proposal of a
novel detector is based on several features of the truly gapless
(armchair) SWNTs. The main property to be utilized is the opening
of the gap in these SWNTs in a magnetic field along the nanotube
axis\cite{DresselhausBook01}. For a $(10,10)$ SWNT this gap is
approximately 1.6~THz in the field  of 10~T. For attainable magnetic 
fields, the gap grows linearly with increasing both magnetic field 
and the nanotube radius. It can be shown\cite{KibisTBP} that 
the same magnetic field also allows dipole optical transitions 
between the top valence subband and the lowest conduction subband, 
which are strictly forbidden in armchair SWNTs without the 
field.\cite{Milosevic03} 

In Figure~\ref{fig6} we show how the energy spectrum and the matrix 
elements of the dipole 
optical transitions polarized along the nanotube axis 
are modified in the presence of a longitudinal magnetic field.  
In the frame of the nearest-neighbor tight binding model, one 
can show that for a $(n,n)$ armchair nanotube the squared 
matrix element of the velocity operator between the states at 
the edge of the gap opened by the magnetic field is given by 
a simple analytic expression:
\begin{equation}
\left|\left\langle\Psi^v_n\left|\hat{v}_z\right|
\Psi^c_i\right\rangle\right|^2
=\frac{4}{3}\left[1-\frac{1}{4}
\cos^2\left(\frac{f}{n}\pi\right)\right]v^2_F,
\label{mag_field_dipole}
\end{equation}
where $f=eBR^2/(2\hbar)$. For experimentally attainable magnetic 
fields, when the magnetic flux through the SWNT is much smaller 
than the flux quantum, the absolute value of the velocity operator 
is close to $v_F$. Equation~(\ref{mag_field_dipole}) is relevant 
to the transitions between the highest valence subband and the 
the lowest conduction subband only for $f\leq1/2$, since for 
the higher values of $f$ the order of the nanotube subbands is 
changed. Notably, the same equation allows to obtain the maximum 
value of the velocity operator in any armchair SWNT for the 
transitions polarized along its axis: this value cannot exceed 
$2 v_F/\sqrt{3}$ (see Fig.~\ref{fig6}c).

\begin{figure}[htb]
\begin{center}
\includegraphics[width=11.5cm]{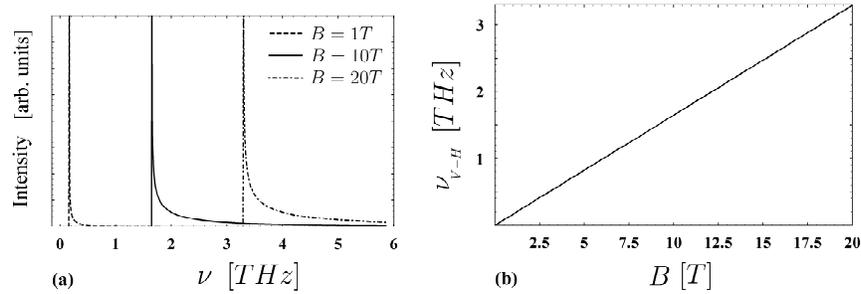}
\end{center}
\caption{(a)~Calculated photon absorption spectra for a 
$\left( 10,10\right)$ SWNT, for three different magnetic field values. 
The absorption intensity is proportional to the product of  
$\left| \left\langle \Psi _{10}^{v}\right| \hat{v}_{z}\left| 
\Psi _{10}^{c}\right\rangle \right|^{2}$ and the joint density
of states. 
(b)~Dependence of the position of the peak in the absorption 
intensity, associated with the Van Hove singularity, on the 
magnetic field.}
\label{fig7}
\end{figure}
The electron (hole) energy spectrum near the bottom (top) of the gap 
produced by the magnetic field is parabolic as a function of a carrier 
momentum along the nanotube axis. This dispersion results in a 
Van Hove singularity in the reduced density of states, which in 
turn leads to a very sharp absorption maximum near the band edge and,
correspondingly, to a very high sensitivity of the photocurrent 
to the photon frequency, see Fig.~\ref{fig7}.

\begin{figure}
\begin{center}
\includegraphics{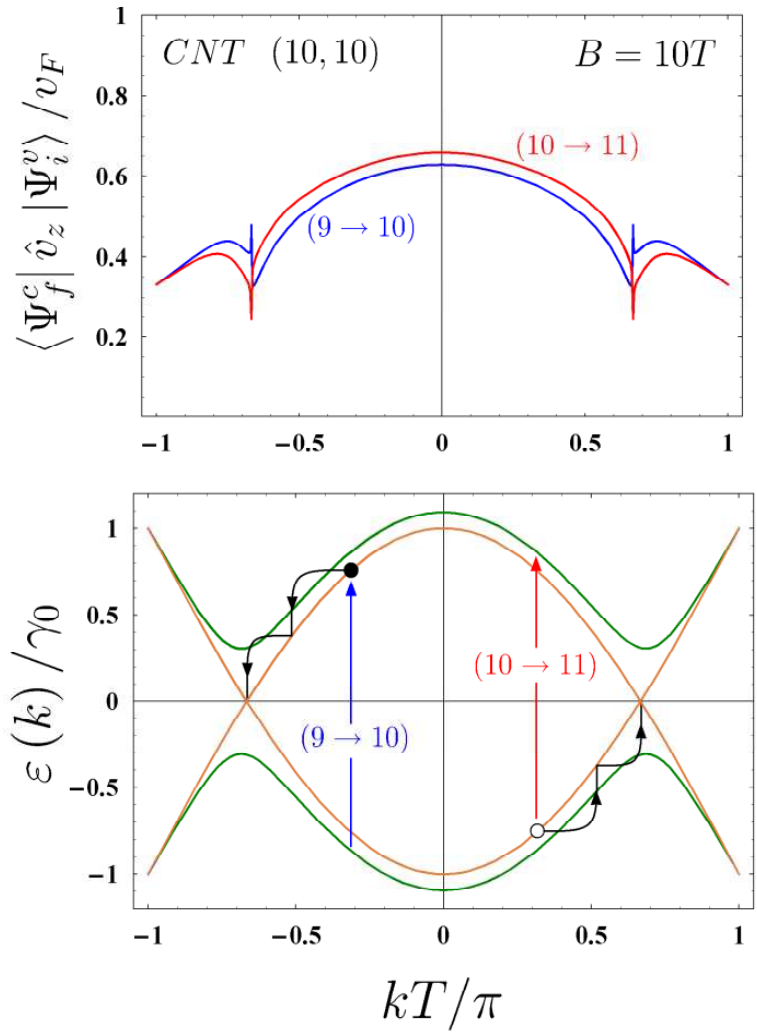}
\end{center}
\vskip 14truecm
\caption{A scheme for creating a population inversion between
the lowest conduction subband and the top valence subband of 
an armchair SWNT in a magnetic field. The upper plot shows the 
calculated matrix elements of the relevant dipole optical 
transitions polarized normally to the axis of a $(10,10)$ SWNT.  
The lower plot shows several energy subbands closest to the 
Fermi level and illustrates the creation of photoexcited 
carriers and their non-radiative thermalization.}  
\label{fig8}
\end{figure}
Notably, the same effect can be used for the generation 
of a very narrow emission line having the peak frequency 
tunable by the applied magnetic field. A population inversion 
can be achieved, for example, by optical pumping with the light 
polarized normally to the nanotube axis, as shown in 
Fig.~\ref{fig8}.

\section{Conclusions}
We have demonstrated that a quasi-metallic carbon nanotube can
emit the THz radiation when the potential difference is applied to
its ends. The typical required voltages and nanotube parameters
are similar to those available in the state-of-the-art transport
experiments. The maximum of the spectral density of emission is
shown to have the strong voltage dependence, which is universal
for all quasi-metallic carbon nanotubes in the ballistic regime.
Therefore, the discussed effect can be used for creating a THz
source with the frequency controlled by the applied voltage.
Appropriately arranged arrays of the nanotubes should be
considered as promising candidates for active elements of
amplifiers and generators of the coherent THz radiation.

We have also shown that an electric field, which is applied
normally to the axis of long-period chiral nanotubes,
significantly modifies their band structure near the edge of the
Brillouin zone. This results in the negative effective mass region
at the energy scale below the high-energy phonon emission
threshold. This effect can be used for an efficient frequency
multiplication in the THz range.

Finally, we have discussed the feasibility of using the effect of
the magnetic field, which opens the energy gaps and allows optical
transitions in armchair nanotubes, for tunable THz detectors and 
emitters.

\acknowledgments
The research was supported by the EU Foundation INTAS (Grants
03-50-4409 and 05-1000008-7801), the Russian Foundation for Basic
Research (Grants 06-02-16005 and 06-02-81012), the Russian
Ministry for Education and Science (Grant RNP.2.1.1.1604), MCT and
FINEP (Brazil). MEP and OVK are grateful to the ICCMP staff for
hospitality.


\begin{thebibliography}{99}
\bibitem{Note_alpha}
For the energy spectrum near the band edge given by 
$\varepsilon=\pm\left[\varepsilon_g^2/4+\hbar^2 v_F^2(k-k_0)^2\right]^{1/2}$, 
it can be shown that $\alpha=\pi/4$.
\end{thebibliography}
\end{document}